\documentstyle[epsfig,12pt,a4]{article}

\begin{document}

\title{
How to quantify deterministic and random influences on the statistics 
of the foreign exchange market
}
\author{
Rudolf Friedrich,\\
Institute f\"ur Theoretische Physik,\\
Universit\"at Stuttgart, D-70550 Stuttgart, \\
Joachim Peinke, Christoph Renner\\
Fachbereich 8 Physik, Universit\"at Oldenburg, D-26111 Oldenburg\\
email r@theo.physik.uni-stuttgart.de; peinke@uni-oldenburg\\}
\maketitle

\begin{abstract}
	It is shown that prize changes of the US dollar - German Mark 
	exchange rates upon different delay times can be regarded as a 
	stochastic  Marcovian process. Furthermore we show that from the 
	empirical data the Kramers-Moyal coefficients can be estimated. 
	Finally, we present an explicite Fokker-Planck equation which models 
	very precisely the empirical probabilitiy distributions. 
\end{abstract}

PACS: 02.50-r;05.10G
\\

Since high-frequency intra-day data are available and easy to access, 
research on the dynamics of financial markets is enjoying a broad 
interest.  \cite{MDal90,MS95,jcV95,HFDF95,MDal95,MS96,BH97}.  
Well-founded quantitative investigations now seem to be feasible.  The 
identification of the underlying process leading to heavy tailed 
probability density function ({\em pdf}) of price changes and the 
volatility clustering (see fig.1) are of special interest.  The shape 
of the pdf expresses an unexpected high probability of large price 
changes on short time scales which is of utmost importance for risk 
analysis.  In a recent work \cite{nature1}, an analogy between the 
short-time dynamics of foreign exchange (FX) market and hydrodynamic 
turbulence has been proposed.  This analogy postulates the existence 
of hierarchical features like a cascade process from large to small 
time scales in the dynamics of prices, similar to the energy cascade 
in turbulence c.f.  \cite{Frisch}.  This postulate has been supported 
by some work \cite{BH97} and questioned by others \cite{MS96}.

One main claim of the hypothesis of cascade processes is that the 
statistics of the time series of the financial market can be 
determined.  The aim of the present paper is to discuss a new kind of 
analysis capable to derive the underlying mathematical model directly 
from the given data.  This method yields an estimation of an effective 
stochastic equation in the form of a Fokker-Planck equation (also 
known as Kolmogorov equation).  The solutions of this equation yields 
the probability distributions with sufficient accuracy (see fig.  1).  
This means that our method is not based on the conventional 
phenomenological comparison between models and several stochastic 
aspects of financial data.  Our approach demonstrates how 
multiplicative noise and deterministic forces interact leading to the 
heavy tailed statistics.  \cite{comment}

Recently it has been demonstrated that from experimental data of 
turbulent flow the hierarchical features induced by the energy cascade 
can be extracted in form of a Fokker-Planck equation for the length 
scale dependence of velocity fluctuations \cite{PRL1}.  In the 
following it will be shown the dynamics of foreign exchange rates can 
also be described by a Fokker-Planck equation.  Here we use a data set 
consisting of 1,472,241 quotes for US dollar-German mark exchange 
rates from the years 1992 and 1993 as used in reference 
\cite{nature1}.

We investigate the statistical dependence of prize changes $\Delta 
x_{i} := x(t+\Delta t_{i})-x(t)$ upon the delay time $\Delta t_{i}$.  
Here $x(t)$ denotes the exchange rate at time t.  We will show that 
the prize changes $\Delta x_1$, $\Delta x_2$ for two delay times 
$\Delta t_1$, $\Delta t_2$ are statistically dependent, provided the 
difference $\Delta t_2-\Delta t_1$ is not too large and that the 
shorter time interval is nested inside the longer one.

In order to characterize the statistical dependency of price changes 
$\Delta x_i$, we have evaluated the joint probability density 
functions
\begin{equation}
  p(\Delta x_2,\Delta t_2;\Delta x_1,\Delta t_1)
\end{equation}
for various time delays $\Delta t_2 < \Delta t_1$ directly from the 
given data set. One example of a contourplot of the logarithms of 
these functions is exhibited in figure 2.  If the two price changes 
$\Delta x_1$, $\Delta x_2$ were statistically independent, the joint 
pdf would factorize into a product of two probability density 
functions:
\begin{equation}
  p(\Delta x_2,\Delta t_2;\Delta x_1,\Delta t_1) =
  p(\Delta x_2,\Delta t_2)p(\Delta x_1,\Delta t_1).
\end{equation}
The tilted form of the probability density $p(\Delta x_2,\Delta 
t_2;\Delta x_1,\Delta t_1)$ clearly shows that such a factorization 
does not hold and that the two price changes are statistically 
dependent.  This dependency is in accordance with observations of 
M{\"u}ller et al.  for cross-correlation functions of the same data 
\cite{MDal95}.  To analyze these correlations in more detail, we 
address the question: What kind of statistical process underlies the 
price changes over a series of nested time delays $\Delta t_i$ of 
decreasing length?  In general, a complete characterization of the 
statistical properties of the data set requires the evaluation of 
joint pdfs $p^N(\Delta x_1,\Delta t_1;....;\Delta x_N,\Delta t_N)$ 
depending on $N$ variables (for arbitrarily large $N$).  In the case 
of a Markov process (a process without memory), an important 
simplification arises: The N-point pdf $p^{N}$ are generated by a 
product of the conditional probabilities $p(\Delta x_{i+1},\Delta 
t_{i+1}|\Delta x_i,\Delta t_i)$, for $i=1,...,N-1$.  As a necessary 
condition, the Chapman-Kolmogorov equation \cite{Risken}
\begin{equation}
\label{chap}
  p(\Delta x_2,\Delta t_2|\Delta x_1,\Delta t_1)=
  \int \hbox{d} (\Delta x_i)\,
  p(\Delta x_2,\Delta t_2|\Delta x_i,\Delta t_i)\, p(\Delta x_i,\Delta
  t_i|\Delta x_1,\Delta t_1) 
\end{equation}
should hold for any value of $\Delta t_i$ embedded in the interval
\begin{equation}
  \Delta t_2 < \Delta t_3 < \Delta t_1 \qquad .
\end{equation}

We checked the validity of the Chapman-Kolmogorov equation for 
different $\Delta t_i$ triplets by comparing the directly evaluated 
conditional probability distributions $p(\Delta x_2,\Delta t_2|\Delta 
x_1,\Delta t_1)$ with the ones calculated ($p_{cal}$) according to 
(\ref{chap}).  In figure 3, the contour lines of the two corresponding 
pdfs are superimposed for the purpose of illustration; the red lines 
corresponding to $p_{cal}$.  Only in the outer regions, there are 
visible deviations probably resulting from a finite resolution of the 
statistics.

As it is well-known, the Chapman-Kolmogorov \cite{Risken} equation 
yields an evolution equation for the change of the distribution 
function $p(\Delta x, \Delta t)$ across the scales $\Delta t$.

For the following it is convenient (and without loss of generality) to 
consider a logarithmic time scale
\begin{equation}
  \tau =-ln \Delta t \qquad .
\end{equation}
Then, the limiting case $\Delta t \rightarrow 0$ corresponds to $\tau 
\rightarrow \infty$.  The Chapman-Kolmogorov equation formulated in 
differential form yields a master equation, which can take the form of 
a Fokker-Planck equation (for a detailed discussion we refer the 
reader to \cite{Risken}):
\begin{equation}\label{KrMoyEnt}
  \frac {d}{d \tau_2} p(\Delta x,\tau)=
  [-\frac{\partial }{\partial \Delta x}
  D^{(1)}(\Delta x, \tau)
  +\frac{\partial^2 }{\partial \Delta x^2} D^{(2)}(\Delta x, \tau)]
  p(\Delta x,\tau)
\end{equation}
The drift and diffusion coefficients $D^{(1)}(\Delta x, \tau)$, 
$D^{(2)}(\Delta x, \tau)$ can be estimated directly from the data as 
moments $M^{(k)}$ of the conditional probability distributions (c.f 
fig.  3):
\begin{eqnarray}\label{MkDk}
  D^{(k)}(\Delta x,\tau) &=& \frac{1}{k!}
  lim_{\Delta \tau \rightarrow 0}  M^{(k)} \\
  M^{(k)} &=& \frac{1}{\Delta \tau}  \int d\Delta x'
  (\Delta x'-\Delta x)^k p(\Delta x', \tau+\Delta \tau|\Delta x, \tau) .
\nonumber 
\end{eqnarray}
As indicated by the functional $\Delta x$ dependency of the moments 
$M^{(k)}$, it turns out that the drift term $D^{(1)}$ is a linear 
function of $\Delta x$, whereas the diffusion term $D^{(2)}$ is a 
function quadratic in $\Delta x$.  In fact, from a careful analysis of 
the data set we obtain the following approximation:
\begin{eqnarray}\label{approx}
  D^{(1)} &=& -0.44 \Delta x \nonumber \\
  D^{(2)} &=& 0.003 \exp{(-\tau/2)}+0.019(\Delta x+0.04)^2 
\end{eqnarray}
($\Delta x$ is measured in units of the standard deviation of $\Delta 
x$ at $\Delta t=40960s$).  With these coefficients we can solve the 
Fokker-Planck equation (\ref{KrMoyEnt}) for the pdf at times $\tau 
>\tau_0$ with a given distribution at $\tau_0$.  Figure 1 shows that 
the solutions of our model nicely fit the experimentally determined 
pdf's \cite{Rem}.  In contrast to the use of phenomenological supposed 
fitting functions c.f \cite{nature1,MS95}, we obtained the changing 
forms of the pdfs by a differential equation.  This method provides 
the evolution of pdfs from large time delays to smaller ones.  This 
definitely is a new quality in describing the hierarchical structure 
of such data sets.  At last, it is important to note that our finding 
of the Fokker-Planck equation for the cascade is in good agreement 
with the previously found phenomenological description in 
\cite{nature1}.  Based on the given exact solution of our 
Fokker-Planck equation \cite{Donkov}, we see that the chosen type of 
fitting function for the pdfs in \cite{nature1} was the correct one.  
Furthermore, we see that the observed quadratic dependency of the 
diffusion term $D^2$ corresponds to the found logarithmic scaling of 
the intermittency parameter in \cite{nature1}, which was taken as an 
essential point to propose the analogy between turbulence and the 
financial market.

We remind the reader that the Fokker-Planck equation is equivalent to 
a Langevin equation of the form (we use the Ito interpretation 
\cite{Risken}):
\begin{equation}\label{stoch}
  \frac{d}{d \tau} \Delta x(\tau)=D^{(1)}(\Delta x(\tau),\tau) +
  \sqrt{D^{(2)}(\Delta x(\tau),\tau)} F(\tau) \qquad .
\end{equation}
Here, $F(\tau)$ is a fluctuating force with gaussian statistics
$\delta$-correlated in $\tau$:
\begin{equation}
  <F(\tau)>=0 \qquad , \qquad <F(\tau)F(\tau')>=\delta(\tau-\tau')
\end{equation}
In our approximation (\ref{approx}) the stochastic process underlying 
the prize changes is a linear stochastic process with multiplicative 
noise, at least for large values of $\Delta x$:
\begin{equation}\label{stoch1}
  \frac{d}{d \tau} \Delta x(\tau)=-0.44 \Delta x(\tau) +
  \sqrt{.019}\Delta x(\tau)  F(\tau) \qquad .
\end{equation}
This stochastic equation yields realizations of prize changes $\Delta 
x(\tau)$, whose ensemble averages can be described by the probability 
distributions $p(\Delta x,\tau)$.  Thus, the Langevin equation 
(\ref{stoch}) produces the possibility to simulate the price cascades 
for time delays from about a day down to several minutes.  
Furthermore, with this last presentation of our results it becomes 
clear that we are able to separate the deterministic and the noisy 
influence on the hierarchical structure of the finance data in terms 
of the coefficients $D^{(1)}$ and $D^{(2)}$, respectively.

Summarizing, it is the concept of a cascade in time hierarchy that 
allowed us to derive the results of the present paper, which in turn 
quantitatively supports the initial concept of an analogy between 
turbulence and financial data.  Furthermore, we have shown that the 
smooth evolution of the pdfs down along the cascade towards smaller 
time delays is caused by a Markov process with multiplicative noise.

Helpful discussions and the careful reading of our manuscript by 
Wolfgang Breymann, Shoaleh Ghashghai and Peter Talkner are 
acknowledged.  The FX data set has been provided by {\em Olsen \& 
Associates} (Z\"urich).

\newpage

\begin{center}
    {\large \bf figure captions}
\end{center}

Figure 1: Probability densities ({\em pdf}) $p(\Delta x, \Delta t)$ of 
the price changes $\Delta x = x(t+\Delta t) - x(t)$ for the time 
delays $\Delta t = 5120, 10240, 20480, 40960 s$ (from bottom to top).  
Symbols: the results obtains from the analysis of middle prices of 
1,472,241 bit-ask quotes for the US dollar-German Mark exchange rates 
from 1 October until 30 September 1993.  Full lines: results form a 
numerical iteration of an effective Fokker-Planck equation with the 
initial condition of the probability distribution for $\Delta t = 
40960 s$.  As drift term $D^{(1)}=- 0.44 \Delta x$ and as diffusion 
term $D^{(2)}= 0.003 exp(-\tau/2) + 0.019(\Delta x + 0.04)^2$ were 
taken.  The units of $\Delta x$ are multiples of the standard 
deviation $\sigma$ of $\Delta x$ with $\Delta t = 40960 s$.  The {\em 
pdf}s are shifted in vertical directions for convenience of 
presentation.  
\\ \\
 
Figure 2: Joint {\em pdf} $p(\Delta x_2,\Delta t_2;\Delta x_1,\Delta t_1)$ for 
the simultaneous occurrence of price differences $\Delta x_1(\Delta 
t_1)$ and $\Delta x_2(\Delta t_2)$ for US dollar-Deutsch mark exchange 
rates.  The contour plot is shown for $\Delta t_1=6168 s$ and $\Delta 
t_2=5120 s$.  $\sigma = 0.0635 DM$ denotes the standard deviation of 
$x(t)$.  The contour lines correspond to $\log p=-1,\ldots,-4$.  If 
the two price changes were statistically independent the joint {\em 
pdf} would factorize into a product of two {\em pdf}s: $p(\Delta 
x_2,\Delta t_2;\Delta x_1,\Delta t_1)= p(\Delta x_2,\Delta 
t_2)p(\Delta x_1,\Delta t_1)$.  The tilted form of the joint {\em pdf} 
provides evidence that such a factorization does not appear for small 
values of $|\log(\Delta t_1/\Delta t_2)|$.  
\\ \\
Figure 3 : Contour plot of the conditional {\em pdf} $p(\Delta 
x_2,\Delta t_2|\Delta x_1,\Delta t_1)=$ $p(\Delta x_2,\Delta 
t_2;\Delta x_1,\Delta t_1)/$ $p(\Delta x_1,\Delta t_1)$ for $\Delta 
t_2= \Delta t_1 / (1.2)^2$ and $t_1=5120$, in the range ($-3\sigma \le 
\Delta x_i \le 3\sigma$, $i=1,2$).  $\sigma$ denotes the standard 
deviation of $x(t)$, see fig.2.  The contour lines correspond to $\log 
p=-1,\ldots,-4$.  For statistically unrelated quantities the 
conditional {\em pdf} would reduce to the probability density 
$p(\Delta x_2,\Delta t_2)$.  \\
In order to verify the Chapman-Kolmogorov equation, the directly 
evaluated {\em pdf} (black lines) is compared with the {\em pdf} 
calculated by $p_{cal}(\Delta x_2,\Delta t_2|\Delta x_1,\Delta 
t_1)=\int dx_i p(\Delta x_2,\Delta t_2|\Delta x_i,\Delta t_i) p(\Delta 
x_i,\Delta t_i|\Delta x_1,\Delta t_1)$ for $\Delta t_2=\Delta t_1 
/(1.2)^2; \Delta t_i=\Delta t_1 /1.2$ (red lines).  Assuming an 
statistical error of the square root of the number of events of each 
bin we find that both {\em pdf}s are statistically identical.

%
%Figure 1
%
\begin{figure}[ht]
  \begin{center}
    \epsfig{file=Fig1.EPSF, width=16.0cm}
  \end{center}
\caption{ }
\end{figure}
%
%
%Figure2
\begin{figure}[htb]
    \begin{center}
       \epsfig{file=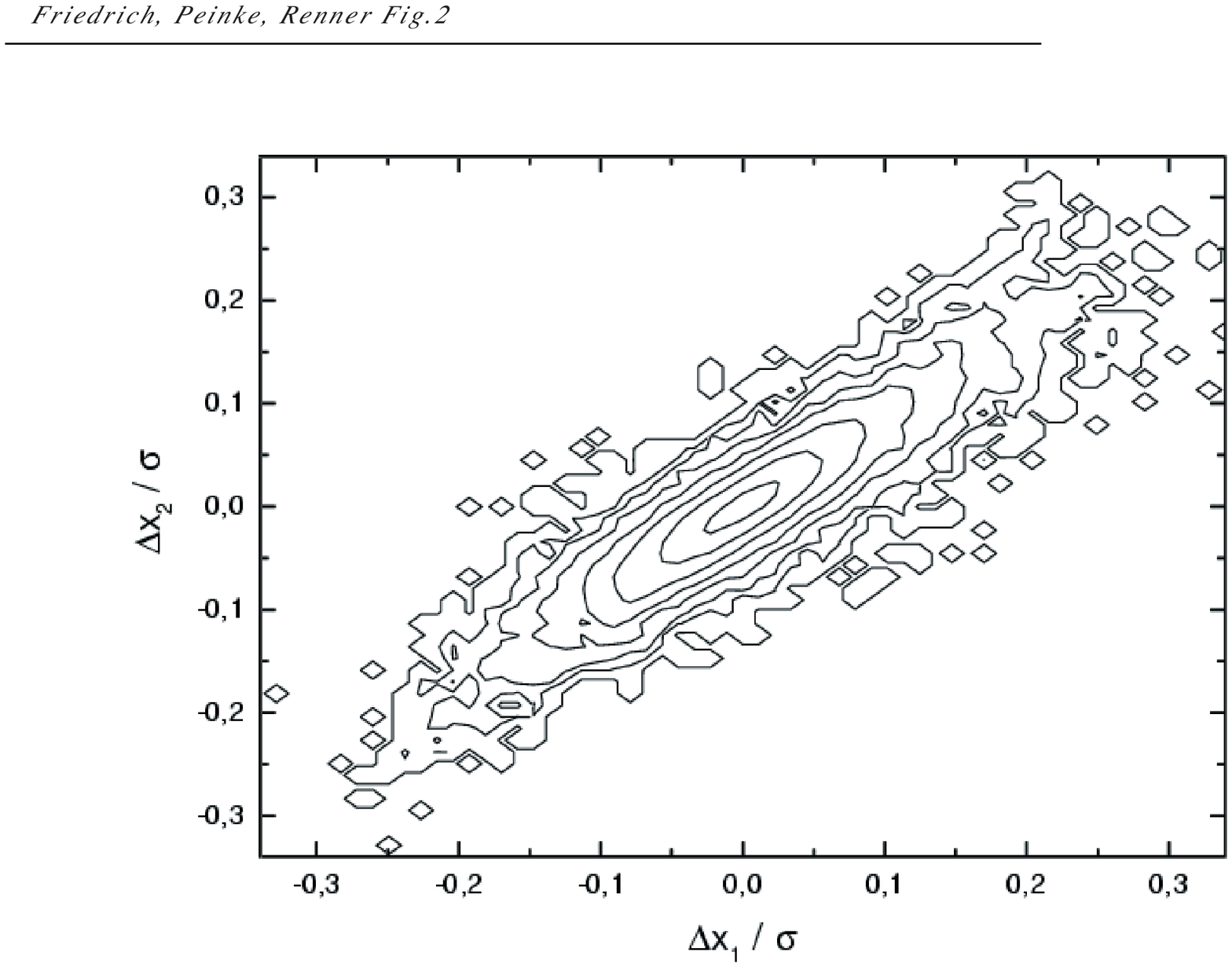,  width=16.0cm }
     \end{center}
\caption{ }
\end{figure}
% 
%
%Figure 3
\begin{figure}[ht]
  \begin{center}
    \epsfig{file=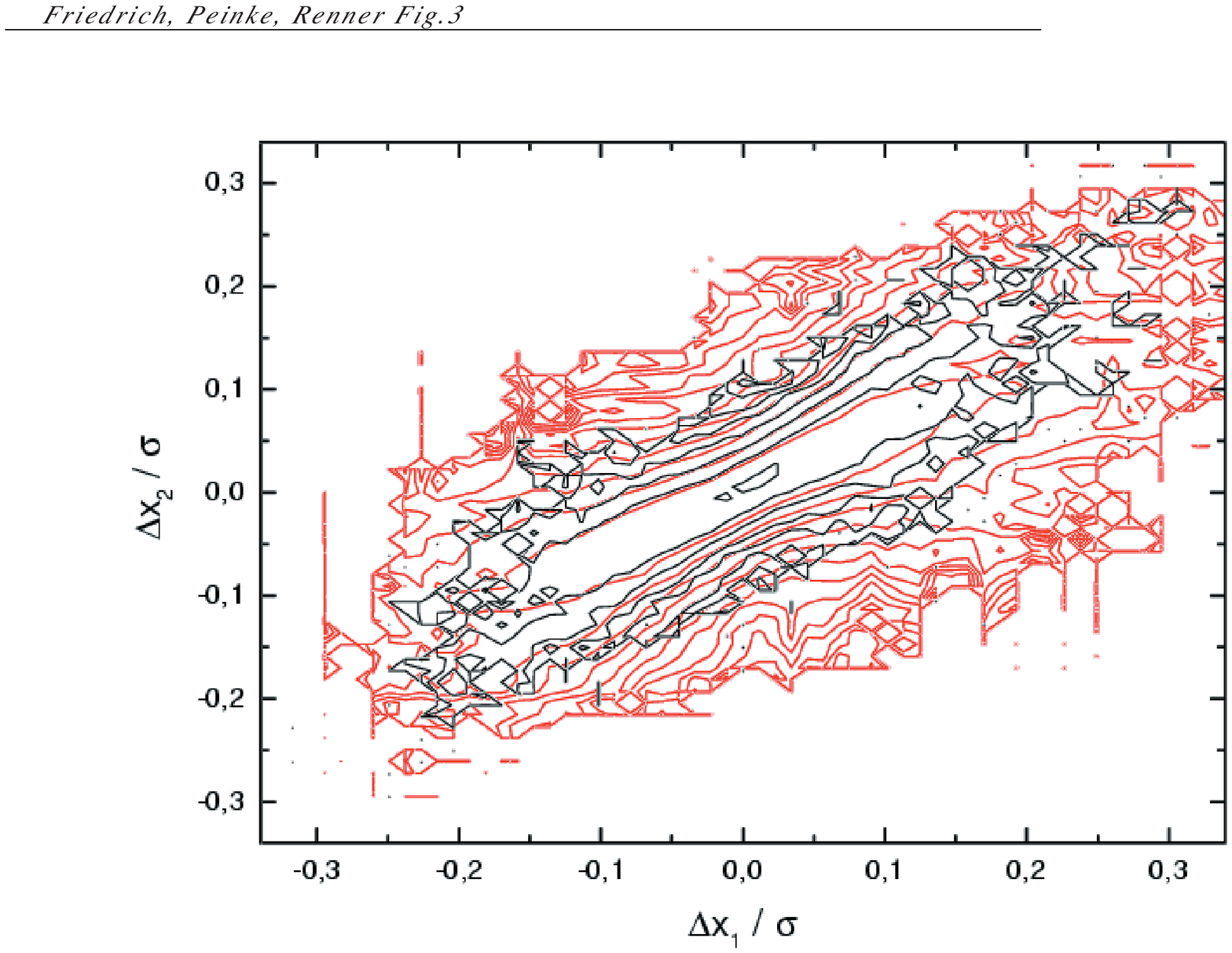, width=16.0cm}
  \end{center}
\caption{ }
\end{figure}
%
%\\


\begin{thebibliography}{10}
\bibitem{comment}
Based on a recent work by Donkov et.al \cite{Donkov} it is easily seen that the 
obtained Fokker-Planck equation is consistent with the 
phenomenological results presented previously \cite{nature1}.
\bibitem{Rem} We remind the reader that
the description of the statistics of price changes by a the Fokker-Planck
equation may only be an approximate one, since the
sample paths need not be continuous functions in time. In this more
general case one is led to an extension of the Fokker-Planck equation by
terms taking care of finite size jumps \cite{Risken}.
\end{thebibliography}
\end{document}